\documentclass[journal=jpccck,manuscript=article]{achemso}

\usepackage[version=3]{mhchem} 
\usepackage[T1]{fontenc}       

\usepackage{subfigure}
\usepackage{verbatim} 
\usepackage{xr} 
\usepackage{color}
\usepackage{lineno, xcolor}

\author{Elsa Perrin}
\affiliation{PASTEUR, D\'epartement de chimie, \'Ecole normale sup\'erieure, PSL Research University, Sorbonne Universit\'es, UPMC Univ. Paris 06, CNRS, 75005 Paris, France}
\alsoaffiliation{Fakult\"at f\"ur Mathematik und Naturwissenschaften,
Stranski-Laboratorium f\"ur Physikalische und Theoretische Chemie, Sekr. C7,
Technische Universit\"at Berlin, Stra\ss e des 17. Juni 135, 10623 Berlin, Germany}
\author{Martin Schoen}
\affiliation{Fakult\"at f\"ur Mathematik und Naturwissenschaften,
Stranski-Laboratorium f\"ur Physikalische und Theoretische Chemie, Sekr. C7,
Technische Universit\"at Berlin, Stra\ss e des 17. Juni 135, 10623 Berlin, Germany}
\alsoaffiliation{Department of Chemical and Biomolecular Engineering, Engineering Building I, Box 7905, North Carolina State University, 911 Partners Way, Raleigh, North Carolina 27695, United States}
\author{Fran\c{c}ois-Xavier Coudert}
\affiliation{Chimie ParisTech, PSL Research University, CNRS, Institut de Recherche de Chimie Paris, 75005 Paris, France}
\author{Anne Boutin}
\affiliation{PASTEUR, D\'epartement de chimie, \'Ecole normale sup\'erieure, PSL Research University, Sorbonne Universit\'es, UPMC Univ. Paris 06, CNRS, 75005 Paris, France}
\email{anne.boutin@ens.fr}

\title{Structure and Dynamics of Solvated Polymers Near a Silica Surface: on the Different Roles Played by Solvent}

\begin{document}

\begin{tocentry}
	\includegraphics[height=3.5cm]{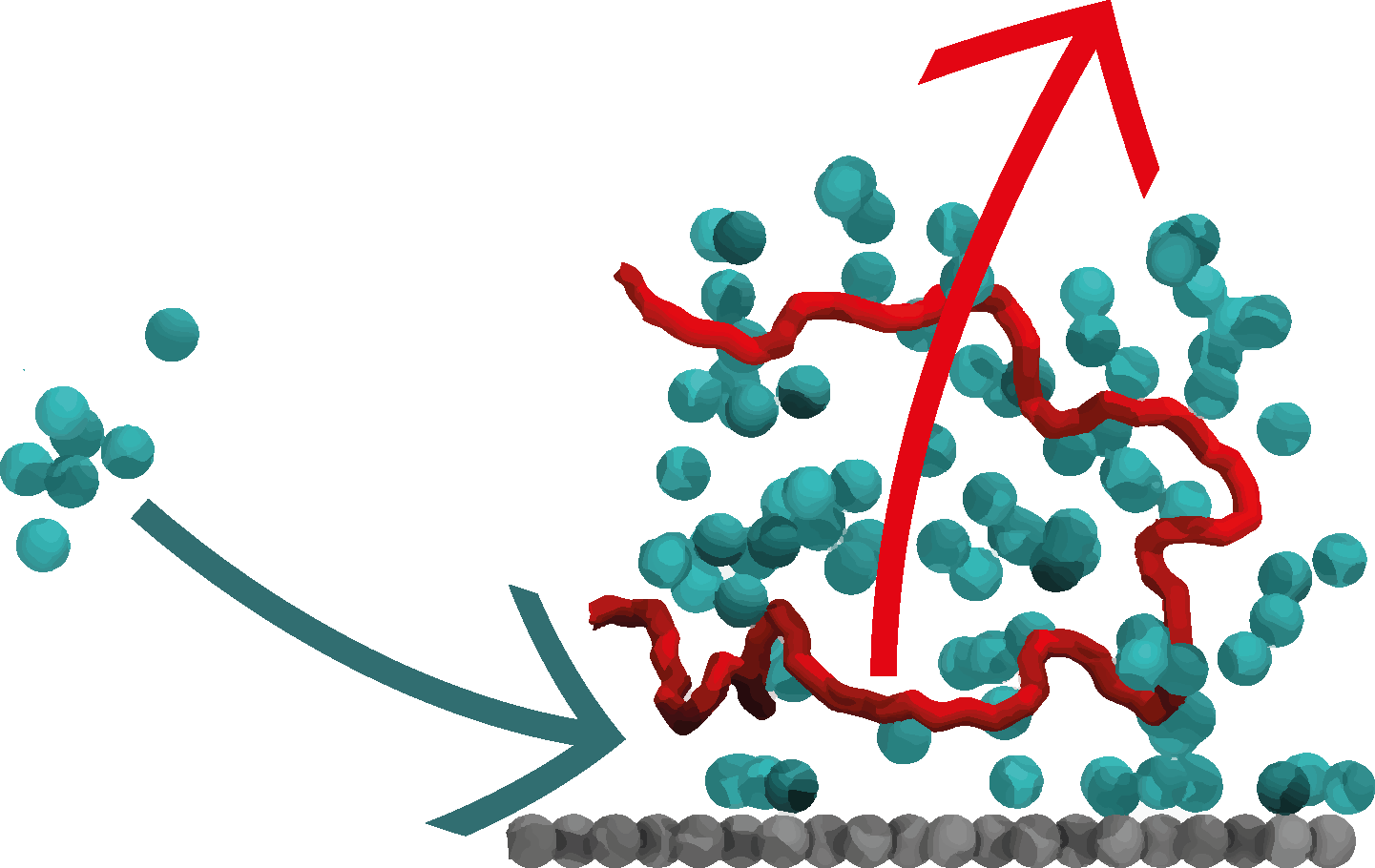}
\end{tocentry}

\begin{abstract}
Whereas it is experimentally known that the inclusion of nanoparticles in hydrogels can lead to a mechanical reinforcement, a detailed molecular understanding of the adhesion mechanism is still lacking. Here we use coarse-grained molecular dynamics simulations to investigate the nature of the interface between silica surfaces and {solvated polymers}. We show how differences in nature of the polymer and the polymer--solvent interactions can lead to drastically different behavior of the polymer--surface adhesion. Comparing explicit and implicit solvent models, we conclude that this effect cannot be fully described in an implicit solvent. We highlight the crucial role of polymer solvation for the adsorption of the {polymer chain} on the silica surface, the significant {dynamic} of fragments of polymer on the surface, and detail the modifications in the structure {solvated polymer} close to the interface.
\end{abstract}

\section{Introduction}\label{lbl:Introduction}
The design of nanocomposite materials is a fast-growing field, both for fundamental research and industrial applications, with many novel and exciting materials such as nanocomposite polymer hydrogels --- reticulated polymer matrices, filled with water ---, or nanoporous materials embedded in an amorphous matrix. Such materials can display an amazing array of mechanical \cite{Sun2012, Sheeney-haj-ichia2002, Mangal2015}, thermal, optical, electrical \cite{Zhao2005}, responsiveness \cite{Liang2000}, frictional \cite{Haraguchi2005} and chemical properties.

Recently the use of nanoparticle solutions as adhesives has been demonstrated as a promising method for gluing soft matter, such as hydrogels or biological tissues \cite{Rose2014, Meddahi-Pelle2014}. Such a method has promising practical applications in the field of surgical adhesives \cite{Annabi2014, Duarte2012, Li2017, Kabiri2011}. At its core, this method relies on adsorption of the polymer onto the surface of the nanoparticles, with the nanoparticles acting as connectors between polymer gels. Local polymer chain rearrangements allow efficient dissipation of energy under stress and retard the fracture of the hydrogel. Although there have been early demonstrations of the feasibility of the method, a clear understanding of the reinforcement mechanisms involving both physical and chemical properties is still lacking and will be important for long-term progress in this area. To better understand the system, Rose and coworkers used hydrogels \cite{Ahmed2015} as a simplified model of biological tissues \cite{Rose2014}. They compared the behavior of two kinds of hydrogels with regard to the surface: polydimethylacrylamide (PDMA) that glues to silica nanoparticles \cite{Hourdet2010} and polyacrylamide (PAAm) which does not adhere to the surface of the silica nanoparticles \cite{Griot1965_1}.

Looking at the chemical structure of the two polymers (Figure~\ref{fig:mapping}), this strong difference in behavior can appear surprising when thinking in terms of hydrogen bonds between silanol terminations (Si--O--H) that are present on the silica surface and in the polymers. PAAm is less sterically hindered than PDMA and one could expect stronger hydrogen bonds with the silica nanoparticles. However PAAm does not adsorb on the silica surface \cite{Griot1965_1}. One explanation could be that isolated silanol groups, {which} are the only binding groups for PAAm \cite{Griot1965_2} (with available hydrogen-bond donnor), are not present on {the surface of the particles} used by Rose and coworkers. Another explanation could rely on the fact that the probability {of forming} hydrogen bonds between a silica surface and the polymer is lower than the probability {of forming} hydrogen bonds between silica and water (due to the high water content of such a hydrogel).
This could explain why PAAm does not glue, however it does not explain why PDMA glues to the silica nanoparticles. Some authors have instead proposed looking at the hydrogel-silica hydrophobic interactions \cite{Doherty2002, Zhang2004}. According to them PDMA adsorbs on the silica surface by hydrogen bonds between oxygen of the dimethylacrylamide group and hydrogen of the silanol group of the silica surface. The ability of PDMA to adsorb on the silica depends on the strength of the hydrogen bond, but also on the hydrophobic interactions of the two alkyl groups from PDMA. The latter reinforces the hydrogen bond through the $p$--$\pi$ conjugation effect between a nitrogen atom and the silanol group and removes water molecules from the silica surface \cite{Zhang2004, Doherty2002, Inomata1990}. This effect does not exist for PAAm because there are no hydrophobic groups. As these hypotheses are not supported by direct experiments, a better understanding of the gluing of a hydrogel to a silica surface can only be acquired by simulation.

Therefore we {investigated} this phenomenon with coarse grained (CG) simulations in order to compute rather large systems at long timescales (50-100~ns). We decided to use a semi-coarse-grained approach in order to keep some of the chemical details. This is a compromise between a detailed atomistic approach and the efficiency of a CG model.
We designed a simple and transferable model between the two polymers in order to focus on the origin of the different behavior between PAAm and PDMA (by limiting the number of parameters).

{Previous CG simulations have concentrated on mechanical properties but were not focused on their chemical origin \cite{Odegard2004, Smith2002}.} The coupling between the local dynamic of the polymer chains on the surface and the resulting mechanical properties is therefore only poorly understood. The solvation of polymer chains, which prevents them from adsorbing on the surface, and the competition between adsorption of solvent or polymer on the surface have, to the best of our knowledge, not been {studied} to date. {We focus on a rather simple system where polymer chains are free of crosslinks and we neglect the curvature of the nanoparticle and consider a flat silica surface.}

The paper is organized as follows: first, we present the CG model employed herein. Then, we use a simple model of a flat silica surface and a polymer solution in implicit water to investigate the local structure and the dynamic of PDMA and PAAm close to the surface. We then introduce an explicit water model to look at the role of the polymer solvation with regard to their adsorption on the silica surface.


\section{Methods}\label{lbl:Methods}
\subsection{Coarse-grained model}\label{lbl:Coarse-grained_methods}

We present here the CG model used for the polymer chain, for the silica surface, and for the water. Coarse-graining a system consists in finding a representation that maps a group of atoms into one CG bead. We designed the CG model within the framework of the Martini force field. The first version of this force field was developed by the Marrink group in 2004 \cite{Marrink2004} and improved in the 2007 version \cite{Marrink2007}, which is the version we use in this work. The overall aim of the Martini CG approach is to provide a simple model that is computationally fast and easy to use, yet flexible enough to be applicable to a large range of biomolecular systems.
The Martini force field has been successfully used for a variety of biomolecules, including lipids \cite{Marrink2004}, sugars \cite{Wohlert2011}, proteins \cite{Monticelli2008}, and polymers such as: polystyrene \cite{Rossi2011}, polyethylene oxide and polyethylene glycol \cite{Lee2009}, polycaprolactam \cite{Milani2011}, and poly(methyl methacrylate) \cite{Uttarwar2012}. The Martini force field is based on a four-to-one mapping, meaning that one CG bead represents on average four heavy atoms and connected hydrogens. This choice is a compromise between computational efficiency and chemical representativeness. The Martini model has four main types of CG bead: polar (P), non-polar (N), apolar (C) and charged (Q). Each particle type has several subtypes. This {allows} for a more detailed description of the underlying atomistic structure. There is a total of eighteen subtypes. For example, polar beads are distinguished by a number indicating the degree of polarity (from 1 for low polarity to 5 for high polarity). Non-polar beads are distinguished by a letter indicating the hydrogen-bonding capability ($d$ for donor and $a$ for an acceptor of a hydrogen bond, $da$ for both and 0 for non). The use of eighteen subtypes of beads allows {one} to set up a rather complicated molecule, with a limited number of building blocks. We {used} them to design the CG model of the polymers PAAm and PDMA, of the silica surface and of the water.

\begin{figure}[!htbp]
  \centering
 \subfigure[CG mapping of the polymers and the silica surface.]{
  \includegraphics[height=5.5cm]{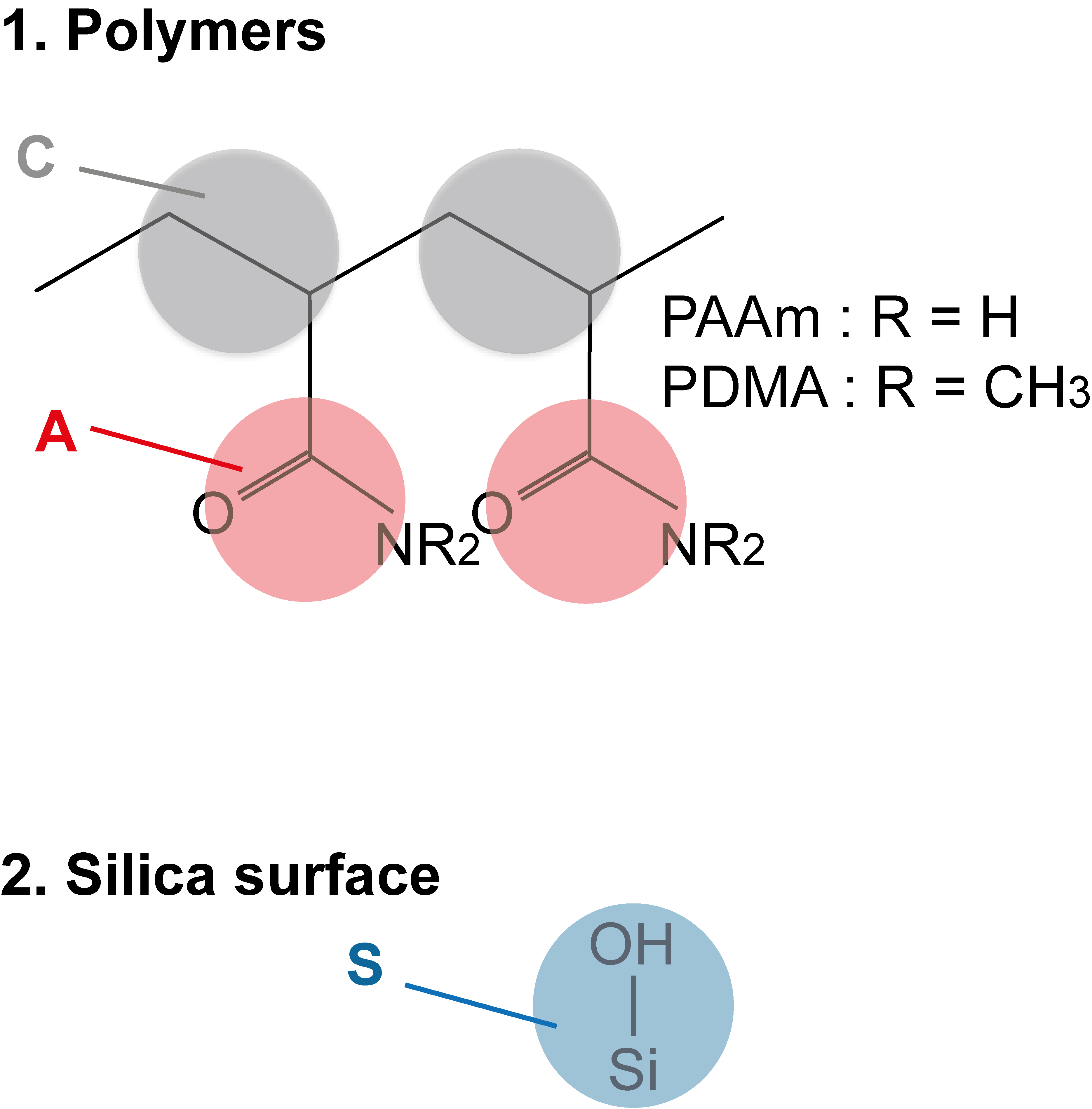}\label{fig:mapping} }
\quad
  \subfigure[Scheme of silanols on the surface of silica.]{
  \includegraphics[height=6cm]{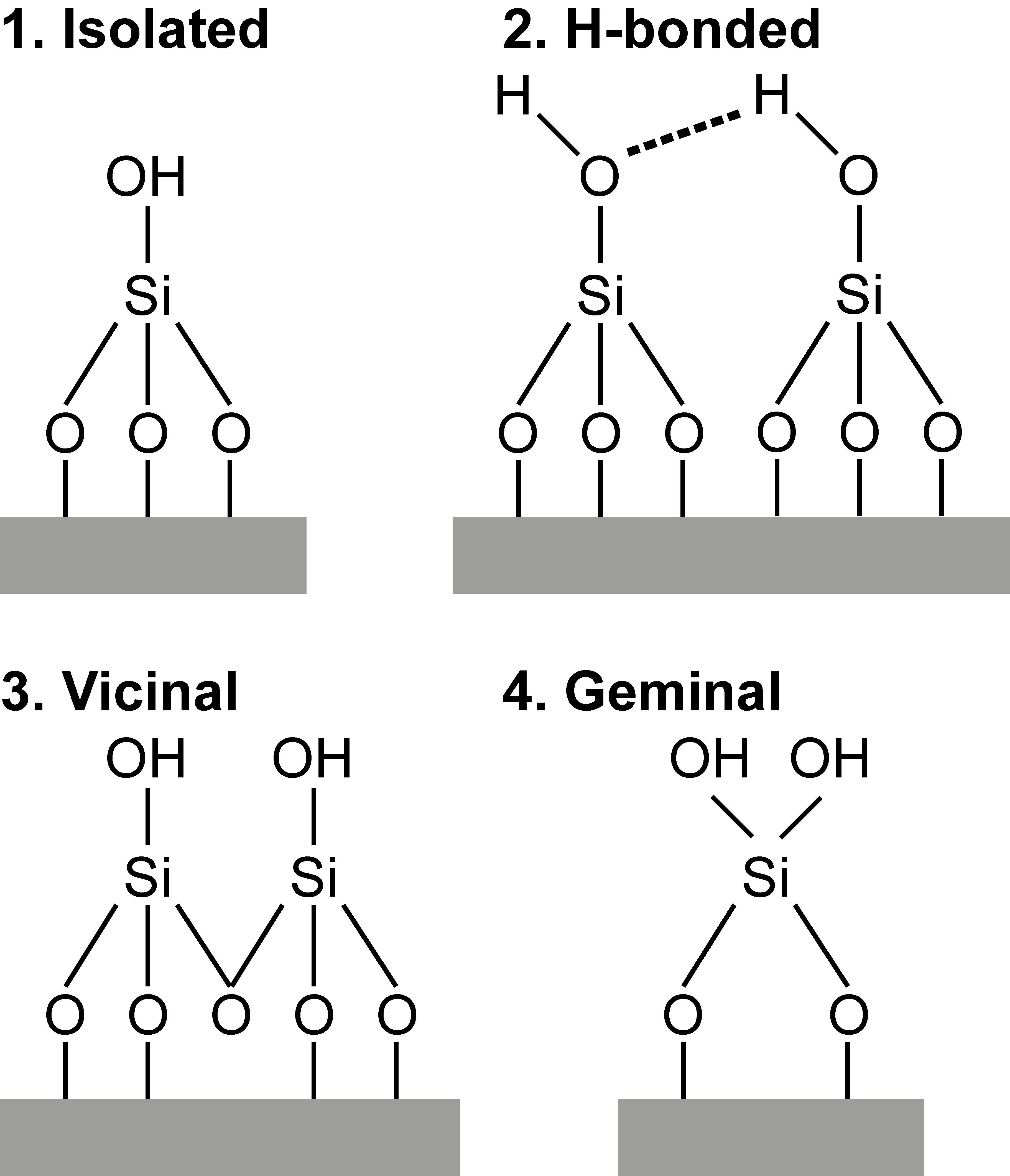}\label{fig:silica} }
 \caption{Left: coarse-grained model of the polymer and the silica surface. Right: scheme of the silanols encountered at the silica surface. }
  \label{fig:mapping+surface}
 \end{figure}

\textit{CG model of PAAm and PDMA. }There are two steps to design a CG model of an all-atoms structure using the Martini force field. The first step is to map chemical groups {onto} CG beads. There are several ways to coarse-grain a system. One of the most widely used is the Kremer-Grest bead-spring model \cite{Kremer1990} where one CG bead corresponds to one or more monomer of the polymer chain. In this description, the CG representation of a polymer is a chain of beads connected with a spring potential. In this work we use a more detailed CG model where one monomer is mapped onto two beads: one bead for the backbone chain (grey circle in figure \ref{fig:mapping}) and one for the side chemical function  (red circle in figure \ref{fig:mapping}). The "C" bead (C for Chain) accounts for two carbon atoms of the backbone chain of PAAm and PDMA. The "A" bead (A for Amide), represents the side groups of PAAm and PDMA. For PAAm, the side group is an amide function. For PDMA, it is a dimethylacrylamide function. Therefore a C bead is similar for PAAm and PDMA, whereas an A bead differs between PAAm and PDMA, {with} an A$_{PDMA}$ bead being bigger than an A$_{PAAm}$ bead. The second step is to associate the corresponding type of bead, using the 18 subtypes we just mentioned. According to Martini's type of particles, a C bead is a bead of type C$_1$: it is an apolar bead with a low degree of polarity. A$_{PAAm}$ is a bead of type N$_{da}$: a non-polar bead which can donate hydrogen bonds (through NH$_2$) or accept hydrogen bonds (through oxygen). A$_{PDMA}$ is a bead of type N$_{a}$ which can only accept hydrogen bonds. Once the type of bead is determined, its size must also be evaluated as the size of the bead plays an important role in the interaction parameters of the system that will be presented later. Martini is based on a four-to-one mapping: beads that map four or more atoms are considered as "normal" beads. Beads that map less than four atoms are "small" beads. C bead is a small bead because it maps only two carbon atoms and connected hydrogens. An A$_{PAAm}$ bead maps three heavy atoms: it is also a small bead. A$_{PDMA}$ is a normal bead because it maps four heavy atoms.

\textit{CG model of the silica surface. }The model we use for the silica surface has a simple geometry and  CG model. First, the geometry of the surface is flat to get rid of the effect of the nanoparticle's curvature at this stage. This choice is motivated by the small curvature of the silica surface in the considered system. The silica nanoparticle we perform matches with silica nanoparticles that are experimentally used \cite{Rose2014}, Ludox TM-50. It is a water solution with concentration of 52 wt\% of silica nanoparticles and radius of 15~nm. Our silica surface has dimensions of $l_x$ = $l_y$ = 63 \AA. Considering a silica nanoparticle with a radius of 15~nm would lead to a height difference of 3 \AA~between the edge of our surface and its center. Such curvature, considering the properties we are interested in (local dynamic quantities, interaction energy, solvation of the polymer chain) can be neglected and a flat silica surface is considered in this work. Silica surfaces exhibit several types of silanol terminations as one can see from figure \ref{fig:silica}) \cite{Rimola2013, Sahai2006}. There are isolated silanols that do not interact with other silanols of the silica surface, H-bonded silanols that interact \textit{via} hydrogen bonds, vicinal silanols which are two silanols forming a siloxane Si--O--Si bridge and geminal silanols wherein one silicon atom bears two hydroxyls. Our CG model is simple in the sense that we choose to map one silanol group into one CG bead called "S" (for silanol). This is shown in the bottom part of the figure \ref{fig:mapping}. We designed our flat surface in order to reproduce the averaged silanol density of an amorphous silica surface, corresponding to 5 OH/nm$^2$ \cite{Zhuravlev1987, Zhuravlev1993, Zhuravlev2000}. Within the Martini framework, silanol beads are non-polar beads (N) which are hydrogen-bond donors, N$_d$. One last thing to note is that the beads of the polymer and of the silica surface are comparable in terms of size. We use a simple surface made of one layer. We checked that using a slab containing several layers (up to eleven layers) does not change the considered properties. Figure \ref{fig:1-11layers} is a comparison of the potential of mean force of PAAm when one or eleven layers are used for the silica surface. They are similar, showing that the use of one silica layer is reasonable.

\textit{CG model of water.} {There are two ways to simulate water within the Martini framework: either explicitly or implicitly.} The Martini model of explicit water comes down to mapping four water molecules {onto} one bead (P$_4$ beads, according to the type of beads) \cite{Marrink2007}.
{The use of the explicit Martini water in a system containing polymer chains has already been done with polystyrene\cite{Rossi2011}, polyethylane oxide and polyethylene glycol\cite{Lee2009}.} The interaction parameter $\epsilon$ between two solvent beads is proposed to be 5.0 kJ.mol$^{-1}$. However, is has been reported that water, modeled as P$_4$ particles, has a freezing temperature that is too high compared to real water. This is linked to the use of a Lennard-Jones 12-6 potential for non-bonded interactions which over-emphasizes the formation of structures \cite{Marrink2013}. This is particularly observed in systems where a nucleation site is already present (like a solid surface in our case). We note, for a system containing a surface and 2064 water beads, a freezing temperature around 360 K. To avoid this problem, we decided to decrease the $\epsilon$ interaction parameter of water beads to 4.5 kJ.mol$^{-1}$. With this new interaction parameter, water close to the surface does not freeze at 300~K (see Supplementary Fig. \ref{fig:water+surface} online). Because we modify the interaction parameter given by Martini for water and we do not strictly use the water from Martini, we decide to use the term "solvent" instead of "water" in our case. However, using an explicit solvent is computationally expensive, especially in our case where {experimental} hydrogels are composed of 90~\% of water in weight. Almost all of the considerable computational time is spent on calculating interactions involving solvent beads.

{One solution is to use implicit solvent. There is a wide range of implicit solvents\cite{Roux1999} employed in polymer science such as Brownian Dynamics (BD)\cite{Carmesin1988}, Dissipative Particles Dynamics (DPD)\cite{Hoogerbrugge1992} or Lattice-Boltzmann\cite{Ahlrichs1999}. In this work, molecular dynamics simulations are performed. We do not account for the random force due to the solvent as in BD or in DPD.} The implicit solvent we use consists of tuning the interaction parameters to take into account the effect of explicit solvent. Indeed, the interactions of a solvophilic bead with other beads would be screened by the presence of a solvation shell in explicit solvent. Therefore, the interaction parameter of solvophilic beads is reduced to take into account the screening due to solvent. On the contrary, there is a depletion of solvent molecules around a solvophobic bead: the interaction parameter of solvophobic beads is increased in an implicit solvent model. We use the implicit solvent CG model developed by Marrink and coworkers called Dry Martini \cite{Arnarez2014}. {This implicit solvent developed within the Martini framework has extensively been used to study systems containing polymers with molecular dynamics \cite{Wang2015_macromolecules, Wang2015, Chong2015, Ginzburg2015, Bochicchio2017}.} Such implicit solvent is easy to use and inexpensive. However we will show that the use of this implicit solvent fails to reproduce details of the microscopic interactions, making the use of explicit solvent necessary in some cases.

\subsection{Nonbonded interactions}\label{lbl:Nonbonded_interactions}
Once the CG model is designed, {the interactions occurring between the beads must be considered.} There are two kinds of interactions: bonded interactions between chemically connected sites and nonbonded interactions. In the Martini force field, non-bonded interactions are described by a Lennard-Jones (LJ) 12-6 potential and there are typically no Coulombic interactions explicitly considered for such noncharged polymers \cite{Rossi2011, Lee2009, Milani2011}.
The strength of the interaction, determined by the value of the LJ welldepth $\epsilon_{ij}$, depends on the interacting beads $i$ and $j$. The $\epsilon_{ij}$ for normal beads (beads that map four {or} more atoms) are found in the interaction matrix given in the paper from Marrink and coworkers \cite{Marrink2007}. For small beads (beads that map two or three atoms), $\epsilon_{ij}$ is scaled to 75\% of the standard value. The effective size of the particles is governed by the LJ parameter $\sigma_{ij}$.  $\sigma_{ij}$, {which is 0.47~nm for normal particles and 0.43~nm for small particles.}
The LJ potential is shifted to zero between $r_{shift} = 0.9~nm$ and $r_{shift} = 1.2$ nm. {Both self and cross} interaction parameters $\epsilon_{ij}$ are given in Table \ref{tab:epsilon}. The upper part of the table summarizes the interaction parameter $\epsilon_{ij}$ for a system containing the explicit solvent we described before. For instance, a C bead interacts with A$_{PAAm}$ bead with an $\epsilon_{ij} = 2.042$ kJ.mol$^{-1}$. But A$_{PAAm}$ and A$_{PDMA}$ do not interact because our system contains eitheir pur A$_{PAAm}$ or pur of A$_{PDMA}$.

\begin{table}
  \caption{Interaction Parameter $\epsilon_{ij}$ (kJ.mol$^{-1}$) Matrix with Explicit and Implicit Solvent.}
  \label{tab:epsilon}
  \begin{tabular}{ c l c l c l c l c l c }
    \hline
     Explicit solvent 	 & C		 & A$_{PAAm}$		  & A$_{PDMA}$ 	& S		 & P$_4$  		\\
    \hline
    C				 & 2.625	& 2.042			 & 2.71			& 2.042	  & 2.3 		\\
    A$_{PAAm}$	 	 & 2.042	& 3.375			 &   ---    			& 3.375	  & 4.5 	  	\\
    A$_{PDMA}$ 	 & 2.71	&  ---     			 & 4.0			& 4.5 	  & 4.5  		\\
    S	 			 & 2.042	& 3.375			 & 4.5			& 3.0 	  & 4.5		 \\
    P$_4$	 	 	 & 2.3	& 4.5 			 & 4.5			& 4.5 	  & 4.5 		\\
    \hline\hline
    Implicit solvent	 & C		 & A$_{PAAm}$		  & A$_{PDMA}$ 	& S		 &   			\\
    \hline
    C			  	 & 3.375	& 2.042			 & 2.71			& 2.042	  &  			\\
    A$_{PAAm}$	  	 & 2.042	& 2.042			 &   ---    			& 2.042	  & 	  		\\
    A$_{PDMA}$  	 & 2.71	&  ---     			 & 2.3			& 2.71 	  &   			\\
    S	 		  	 & 2.042	& 2.042			 & 2.71			& 1.725 	  & 			 \\
  \end{tabular}
\end{table}

\subsection{Bonded interactions}\label{lbl:Bonded_interactions}
Bonded interactions between two chemically connected beads are described by a harmonic potential $V_{bond}(r)$:

\begin{equation}
	V_{bond}(r) = \frac{1}{2}K_{bond}(r-r_0)^2,
	\label{eq:Vbond}
\end{equation}
where $r$ is the distance between two beads, $K_{bond}$ is the force constant of the harmonic bonding potential, and $r_0$ is the equilibrium distance. Angles between three neighboring beads are governed by the potential $V_{angle}(\theta)$:

\begin{equation}
	V_{angle}(\theta) = \frac{1}{2}K_{angle}[cos(\theta)-cos(\theta_0)]^2,
	\label{eq:Vangle}
\end{equation}
where $\theta$ is the angle between the three beads, $K_{angle}$ is the force constant and $\theta_0$ is the equilibrium angle. We decided not to use dihedral potential for the sake of simplicity. The Martini paper \cite{Marrink2007} provides general values for $K_{bond}$, $K_{angle}$, $r_0$ and  $\theta_0$. We find those values not adapted to our system, which is a rather fine grained model, $i.e$ containing small beads. We have thus reparametrized our CG model, based on input from all-atoms simulations. Parameters for the bond and angle potentials were obtained by comparing distributions from all-atom simulations with distributions from CG simulation.

\begin{figure}[!htbp]
  \centering
 \subfigure{
  \includegraphics[height=5cm]{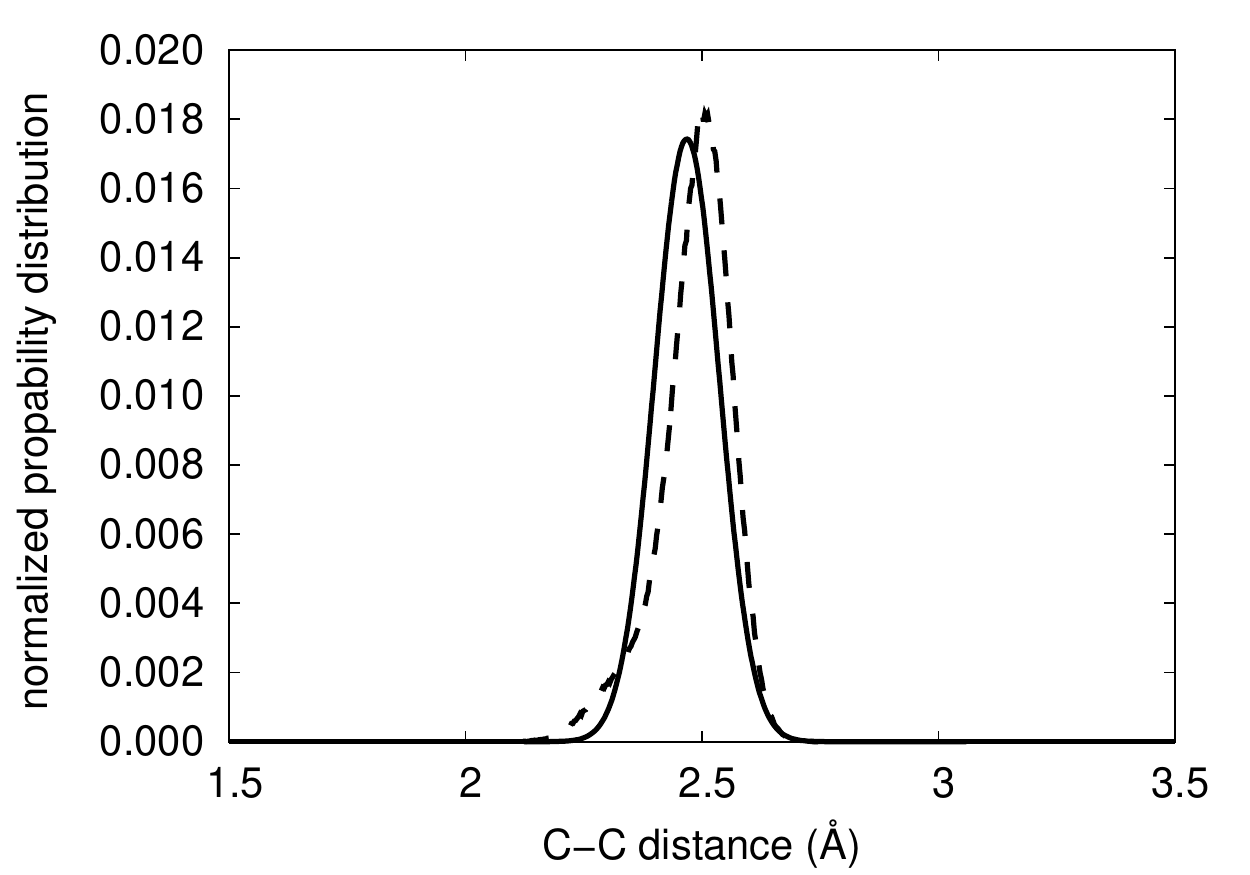} }
\quad
  \subfigure{
  \includegraphics[height=5cm]{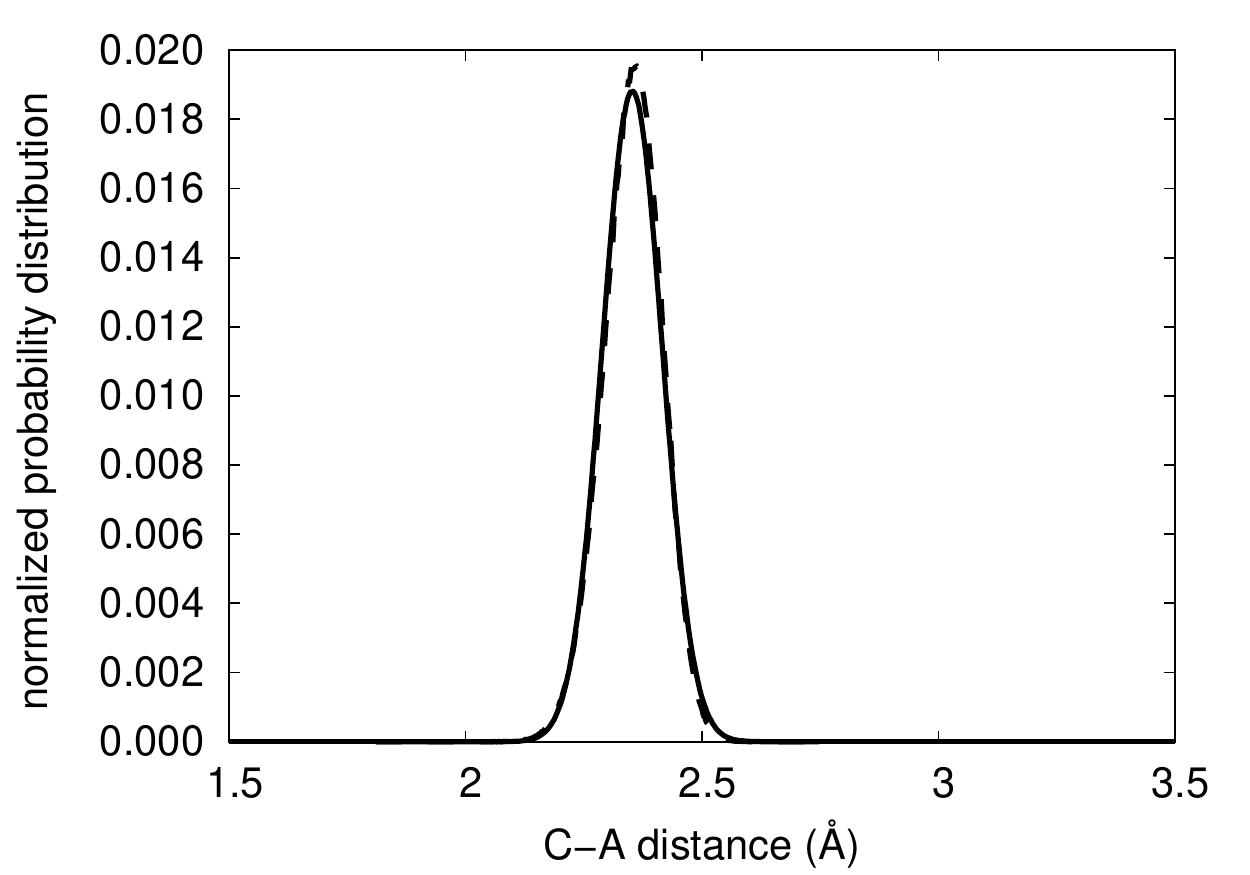} }
\quad
  \subfigure{
  \includegraphics[height=5cm]{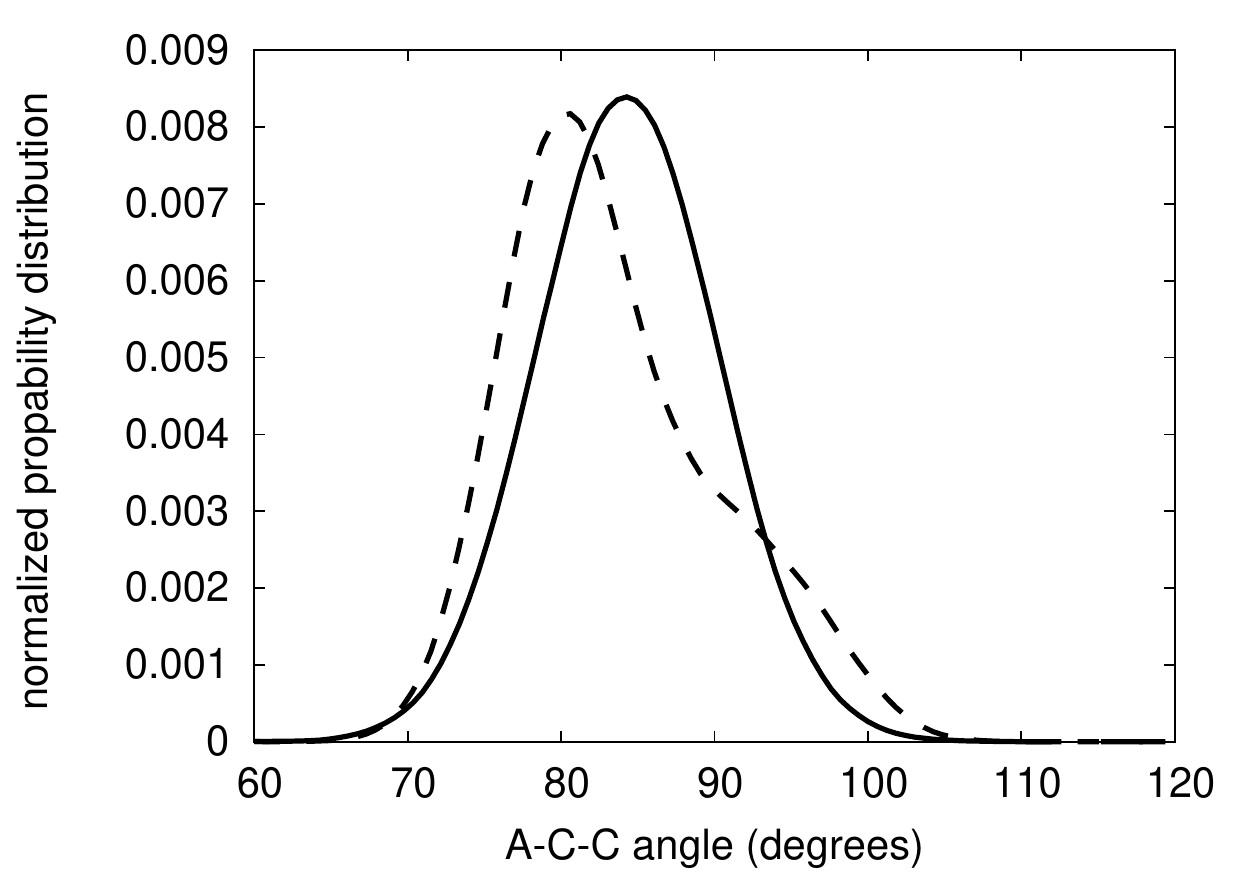} }
\quad
  \subfigure{
  \includegraphics[height=5cm]{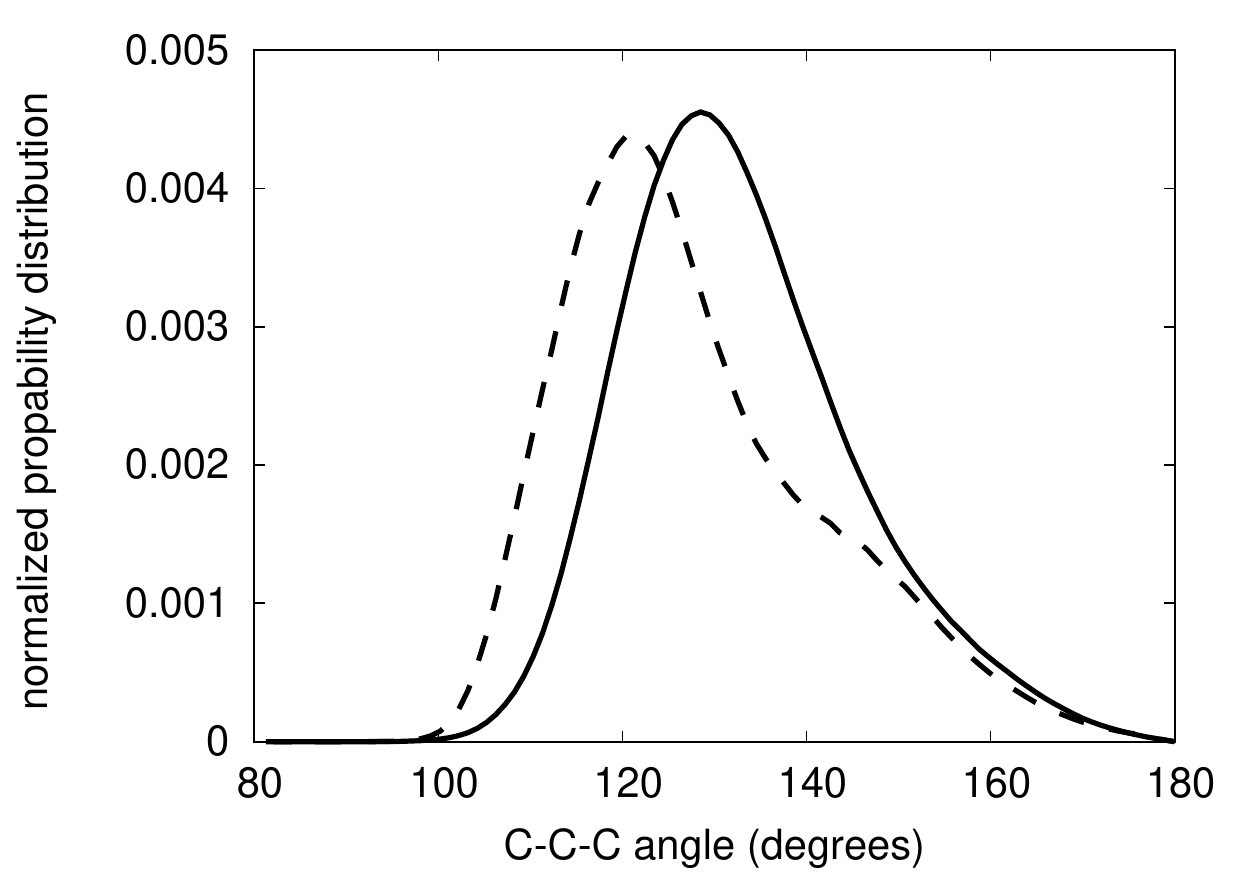} }
 \caption{Comparison between all atoms (dashed line) and coarse-grained (plain line) bonds and angles distribution. }
  \label{fig:AA+CG}
 \end{figure}

We ran all-atoms simulations, using the LAMMPS code \cite{Plimpton1995}, \cite{LAMMPS} and the CHARMM22 force field \cite{MacKerell1998} (the timestep {was} set to 1 fs). The CHARMM force field is widely used for all atoms simulations and is adapted to the system we study. Along the all-atoms simulation we compute lengths and angles between the center of mass of the groups of atoms that are mapped onto one CG bead. The resulting histograms of bond length $r$ and of the angle $\theta$ are the red curves in figure \ref{fig:AA+CG}. The CG histograms are the green curves in \ref{fig:AA+CG}. Both histograms are normalized. They indicate the probability {of finding} a bond at a given distance or an angle at a given value.

All-atoms and coarse-grained simulations are done with a system containing five polymer chains of twenty monomers. The averaged value of the all-atoms histograms is used to optimize the equilibrium bond length $r_0$ and angle $\theta_0$ of the CG potentials $V_{bond}(r)$ (Eq. \ref{eq:Vbond}) and $V_{angle}(\theta)$ (Eq. \ref{eq:Vangle}). We tune $r_0$ and $\theta_0$ in such a way {so} that the averaged value of the CG histogram matches the averaged value of the all-atoms histogram. The coupling constants $K_{bond}$  and $K_{angle}$ are optimized by comparing the width of the CG histogram with the corresponding width of the all-atoms histogram. End-to-end distances (the distance between the two extremities of a polymer chain) and radius of gyration between all-atoms and CG simulation are finally compared to validate our CG model. We find, for the radius of gyration, an average value of 7.1 $\pm$ 0.3 \AA~for the all-atoms system and of 7.3 $\pm$ 1.0 \AA~for the coarse-grained system. The values of the radius of gyration are in very good agreement between all-atoms and coarse-grained systems. As for the end-to-end distance, the average value is 15.07 $\pm$ 2.53 \AA~in the all-atoms system and 13.21 $\pm$ 3.33 \AA~in the coarse-grained system. End-to-end distances also show good agreement between all-atoms and coarse-grained simulations. The final K$_{bond}$, K$_{angle}$, r$_0$ and  $\theta_0$ parameters are summarized in Table \ref{tab:bonded}. For the sake of simplicity and to have a CG model that differs as little as possible between PAAm and PDMA, K$_{bond}$, K$_{angle}$ and $\theta_0$ are the same for PAAm and PDMA. Only A$_{PDMA}$--C differs. This arises from the comparison between all-atoms and CG distributions.

\begin{table}
  \caption{Bonded Interaction Parameters for Bonds and Angles, for PAAm and PDMA. }
  \label{tab:bonded}
  \begin{tabular}{ c l c l c }
    \hline
Bonds		 	 & K$_{bond}$ (kJ.mol$^{-1}$.nm$^2$)		& r$_0$ (nm)  		\\
    \hline
C--C 			& 566.7								& 0.249				\\
A$_{PAAm}$--C	& 666.7								& 0.237				\\
A$_{PDMA}$--C	& 666.7								& 0.271				\\
    \hline\hline
Angles     			 & K$_{angle}$ (kJ.mol$^{-1}$)				& $\theta_0$ (degrees)  		\\
    \hline
C--C--C  			&116.7								& 127.5				\\
A$_{PAAm}$--C--C	& 233.3 								& 85.5				\\
A$_{PDMA}$--C--C	& 233.3 								& 85.5				\\
  \end{tabular}
\end{table}

\subsection{Numerical details}\label{lbl:Numerical_details}

CG molecular dynamic simulations were done using the LAMMPS simulation package \cite{LAMMPS, Plimpton1995} and the Martini force field developed by Marrink and coworkers \cite{Marrink2004, Marrink2007,Marrink2013, Arnarez2014}. The LJ interactions were smoothly shifted to zero between 0.9 and 1.2 nm by using the following formula:
\begin{align}
E {}& =  4 \epsilon \bigg[ \Big( \frac{\sigma}{r} \Big) ^{12} - \Big( \frac{\sigma}{r} \Big) ^{6}  \bigg] \qquad \qquad {}& \text{if} \quad r < r_{in}  \\
 {}& = 4 \epsilon \bigg[ \Big( \frac{\sigma}{r} \Big) ^{12} - \Big( \frac{\sigma}{r} \Big) ^{6}  \bigg]
\frac{[r^2_{out} - r^2]^2[r^2_{out} + 2r^2 - 3r^2_{in}]}{[r^2_{out} - r^2_{in}]^3}
 \qquad \qquad {}& \text{if} \quad r_{in} < r < r_{out} \nonumber \\
  {}& = 0 \nonumber
 \qquad \qquad {}& \text{if} \quad  r > r_{out},
\end{align}
where $r_{in}$ = 0.9~nm and $r_{out}$ = 1.2~nm and $r$ is the distance between two particles. The timestep was set to 10 fs and coordinates were saved every 5 ps for analysis. Periodic boundary conditions in $x, y, z$ directions were used. The systems were first equilibrated in the NPT ensemble and then in the NVT ensemble during 5~ns at a pressure of 1 atm and a temperature of 300~K. For both systems, simulations are done in the NVT ensemble. Production simulations were performed in the NVT ensemble during 50~ns to 100~ns. Nose-Hoover thermostat and barostat were employed and the pressure was set to one atmosphere. The relaxation time was set to 100 fs for the temperature and to 1000 fs for the pressure. Several independent simulations (5 or 10) are performed and are used to compute averaged quantities. Input files for system with an implicit or with an explicit solvent are provided in the Supplementary material. No coupling method was used for the interaction parameters.


\section{Results and discussion}\label{lbl:Results_and_discussion}
\subsection{Solvated polymer interacting through an implicit solvent on a surface}
\label{lbl:polymelt}

\begin{figure}[!htbp]
  \centering
   \subfigure[ Side view of three PDMA chains (in orange, blue and gray) on silica surface (black beads).]{
  \includegraphics[height=6cm]{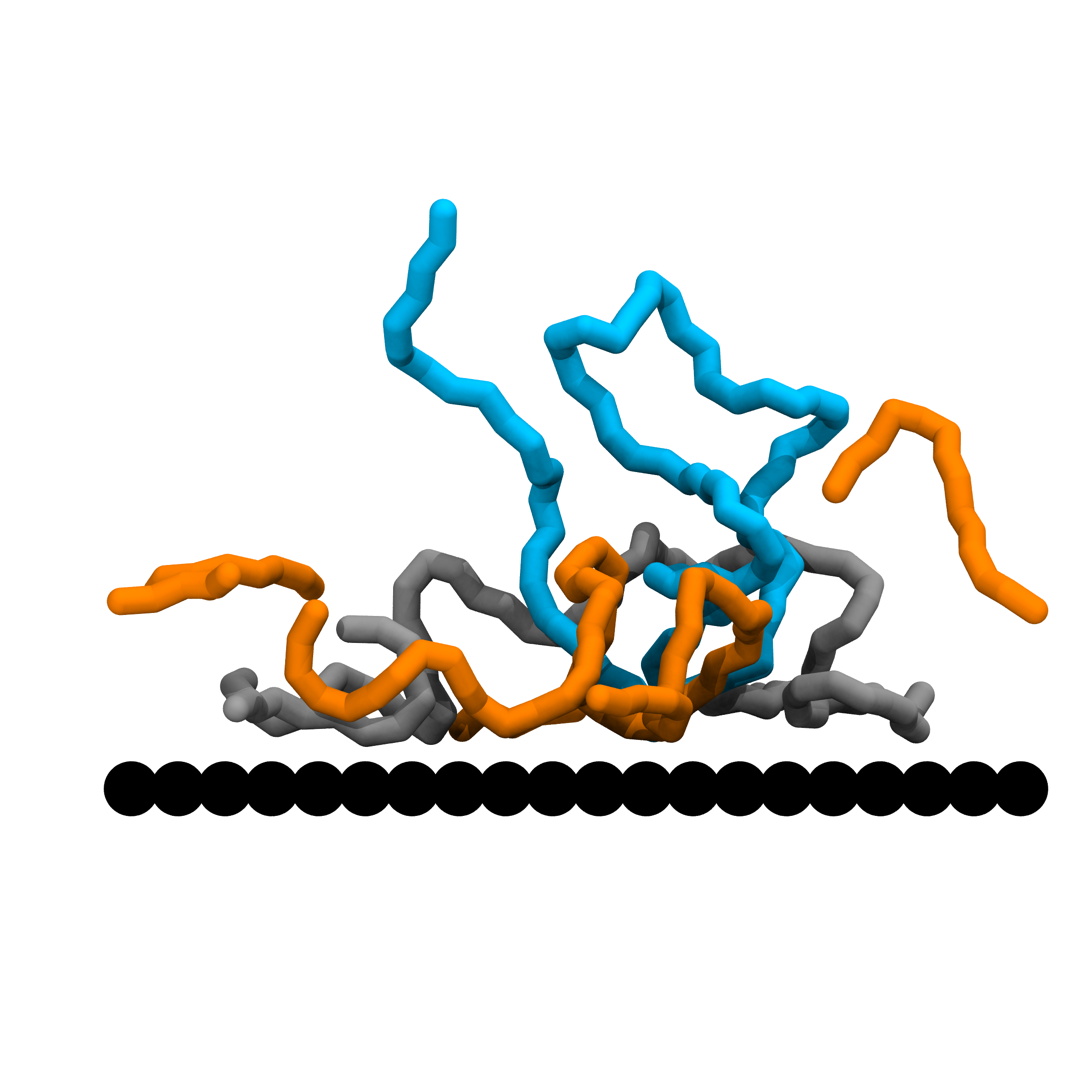}\label{fig:PDMA+surface_melt_side} }
\quad
  \subfigure[Top view of A$_{PDMA}$ beads (red beads) on silica surface (black beads) in region I.]{
  \includegraphics[height=5.5cm]{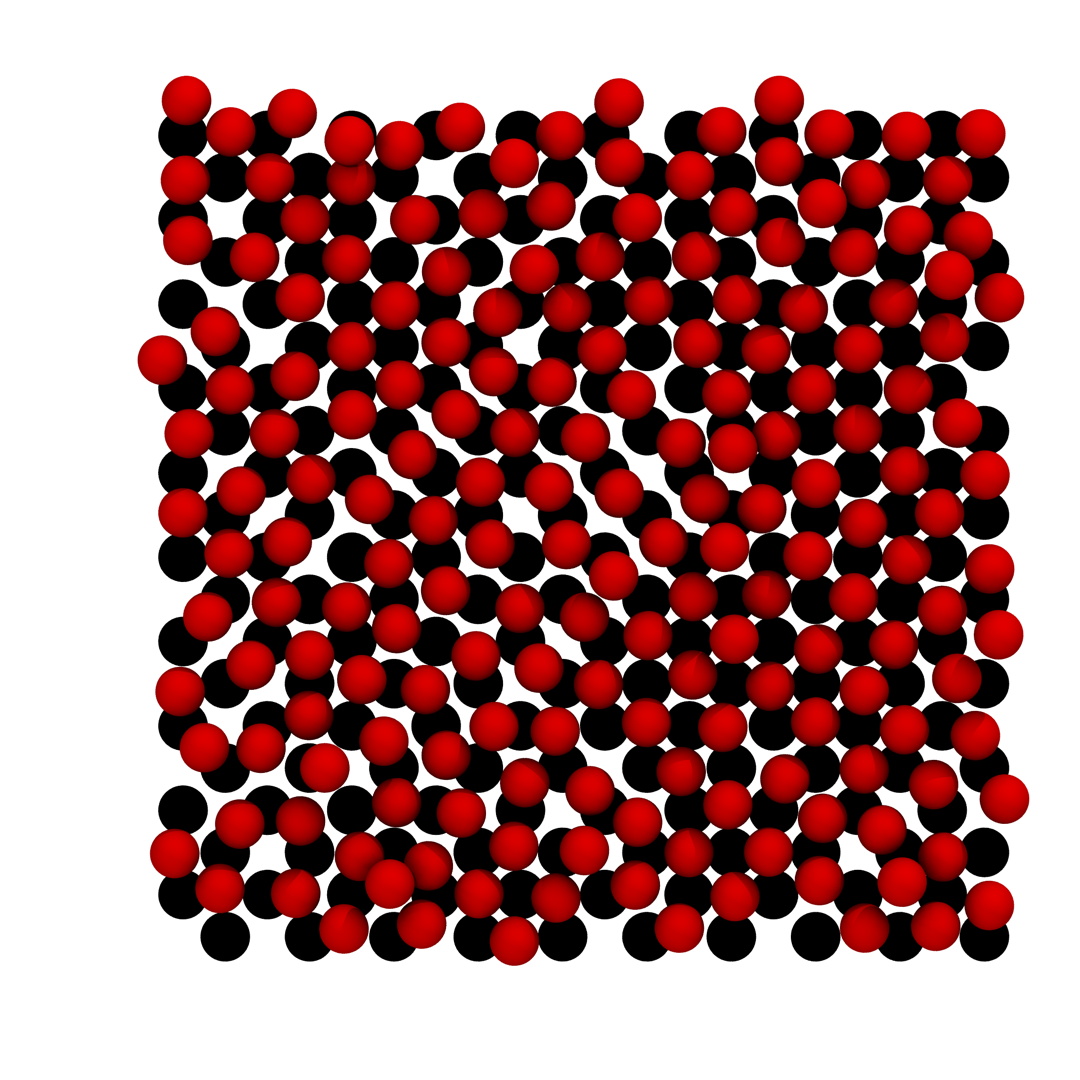}\label{fig:PDMA+surface_melt_top} }
 \caption{Top and side view of the polymer conformation on silica surface.}
  \label{fig:PDMA+surface_melt}
 \end{figure}

To understand the influence of the nanoparticles on the polymer network, we first look at the local structure and dynamics at the interface between the surface {of the} nanoparticles and the polymers. As a starting point, we {studied} a polymer {solution} near a surface interacting through an implicit solvent. The system is composed of a flat surface (200 S beads) and 24 chains of 90 monomers (2160 A beads and 2160 C beads). {A snapshot of the simulation box is displayed on Fig. \ref{fig:snapshotPAM_nolength} of the Supplementary material. The simulation box dimensions are : $l_x$, $l_y$ = 63 \AA, $l_z$ = 51.4 \AA~for PAAm and $l_z$ = 55.8 \AA~for PDMA after equilibration at a pressure of 1 atm.} We checked that the polymer slab is large enough so that polymer beads located in the middle of the slab do not feel any interaction from the surface. {The system was first equilibrated during 5~ns in the NPT ensemble at a pressure of one atmosphere ({pressure was applied in the direction normal to the surface}) and a temperature of 300~K, then during 5~ns in the NVT ensemble.} We follow the dynamic for 50~ns and save the coordinates of the beads along the simulation. The results we present here are averaged over 5 independent simulations of 50~ns each. As A beads interact more strongly with the surface than C beads and are different between PAAm and PDMA, we concentrate only on A beads for the rest of this work. Figure \ref{fig:PDMA+surface_melt_side} is a {side} view of the polymer conformation on the silica surface where only three over twenty-four PDMA chains are represented. It illustrates how polymer chains are well entangled and form trains and loops close to the silica surface. {Figure \ref{fig:PDMA+surface_melt_top} shows that A beads in region I are well dispersed on the silica surface.}

\begin{figure}[!htbp]
  \centering
	\includegraphics[height=8cm]{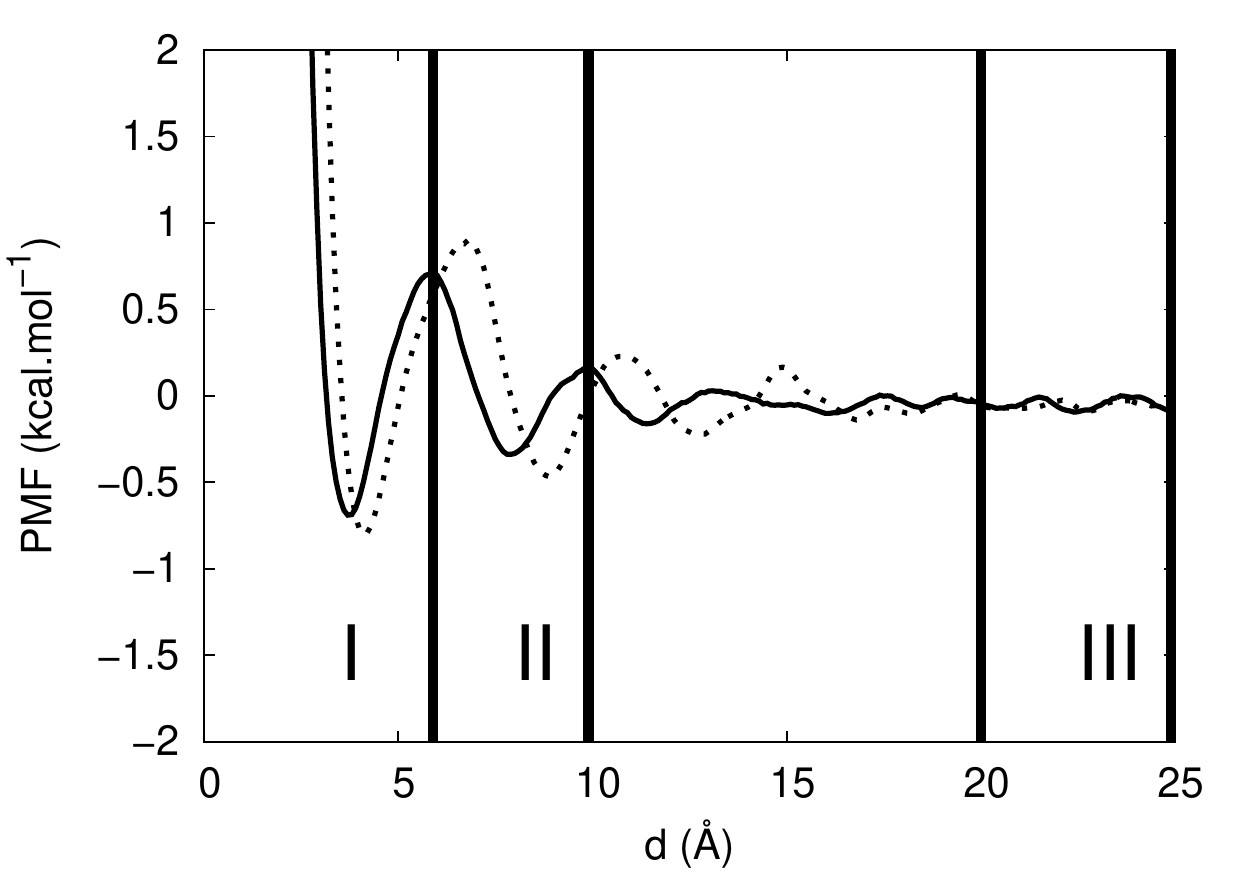}
   \caption{Potential of mean force of PAAm (plain line) and of PDMA (dashed line). Implicit solvent.  }
  \label{fig:PMF_PAAm+PDMA+surface}
\end{figure}

We computed the normalized histogram $P(d)$ of the distance $d$ between the surface and the polymer A beads. The free energy profile of the A beads is then $F(d) = -k_BTln[P(d)]$, plotted in figure \ref{fig:PMF_PAAm+PDMA+surface}.The surface introduces a symmetry breaking that forces the polymer beads to organize in layers. At small $d$, there is a well corresponding to a high stability of the polymer beads: they form a first layer in the direct vicinity of the surface, then a second layer. Beyond that second layer, there is no more influence of the surface on the polymer beads. Both PAAm and PDMA beads are similarly structured close to the surface.

From this structure we define different regions of stability that are represented in figure \ref{fig:PMF_PAAm+PDMA+surface}. Region I is defined as the first layer near the surface: it starts close to the surface and ends at the first maximum as one can see from figure \ref{fig:PMF_PAAm+PDMA+surface}. A beads are well dispersed on the silica surface, as can be seen in Fig. \ref{fig:PDMA+surface_melt_top}. Region II starts just after region I and ends at the second peak of the probability distribution. This second layer is less structured than the first layer. Region III corresponds to a bulk-like region where there is no interaction between the polymer and the surface and is defined by a slab of 10~{\AA} in the middle of the box.

From an energetic point of view, in region I, the averaged interaction energy between the surface and A beads is 0.9 kcal.mol$^{-1}$ per bead for PAAm and 1.6 kcal.mol$^{-1}$ per bead for PDMA. It is consistent with the interaction parameters that are given in Table \ref{tab:epsilon}: A$_{PDMA}$ beads interact more strongly with S beads than A$_{PAAm}$. This is interesting to compare those values with the free energy barrier $\Delta F_B$, obtained by subtracting the value of the PMF at the first peak from the PMF's value at the first well. It characterizes the local exchange of the A beads, or the ability to detach/attach an A bead from the surface. $\Delta F_B$ turns out to be slightly higher for PDMA (1.7 kcal.mol$^{-1}$) than for PAAm (1.4 kcal.mol$^{-1}$). It is thus a little harder for an A bead of PDMA to leave the first adsorption layer to go to the second layer or to the bulk. The results obtained for the free-energy barrier and for the interaction energy between A beads and the surface in region I are then consistent. We can conclude that there is a slight difference between PAAm and PDMA when the behavior in the vicinity of the surface is considered. A beads of PDMA interact more strongly with the surface, making it more difficult for them to leave the first adsorption layer.

Our aim is now to go beyond energetic and thermodynamic information and to study whether they have consequences on the ability of the polymer beads to move in the vicinity of the surface or not. Indeed, the capacity of polymer beads to attach and detach and to reorganize near the surface is linked to the mechanical properties of the resulting nanocomposite system. If the system has the ability to rearrange in an efficient way, it will dissipate energy under stress and will lead to a stronger system \cite{Montarnal2011, Carlsson2010}.
In order to quantify the ability of the polymer beads to move close to the surface, we consider as "active beads" A beads that cross regions over the course of the simulation (regions being defined in Fig. \ref{fig:PMF_PAAm+PDMA+surface}).
Polymer beads undergo two kinds of events: whether they leave region I (close to the surface) or whether they reach region I. Beads that leave region I go from region I to beyond region II. Beads that reach region I arrive in region I from outside region II. The averaged number of events is in the second row of Table \ref{tab:activity_A_hydrogel}.
The number of occurrences of the second event divided by the simulation time is the frequency of this event and is in the third row of the table. We compare the numbers between A beads of PAAm and of PDMA.

\begin{table}
  \caption{Dynamical quantities of PAAm and PDMA's A Beads Interacting through an Implicit Solvent.}
  \label{tab:activity_A_hydrogel}
  \begin{tabular}{ c l c l c }
    \hline
Dynamic									& A beads of PAAm 	& A beads of PDMA \\ \hline
Number of active beads 						& 38.8($\pm$12.3) 	& 21.4($\pm$13.1)  \\
Averaged number of events per active bead		&  1.9($\pm$0.4)	& 3.8($\pm$2.0)  \\
Frequency of the event "bead leave" (Ghz)    		&  1.0			& 0.8
\end{tabular}
\end{table}

First of all, it is worth noting that PAAm and PDMA have rather few active beads over the course of the simulations. In order to be sure that active beads are not only end-of-the-chain beads but are rather well dispersed along the polymeric chain, we checked where {the active beads are located} (see Supplementary Fig. \ref{fig:active_beads} online). Figure \ref{fig:active_beads} shows the position of active beads along doubled up polymer chains, averaged over the 24 chains and over the course of the simulations: position 1 is the end of the chains and position 45 is right in the middle of the chain. The figure shows an excess of active beads at the end of the chains, but also active beads all along the chains. Therefore we are confident that the {dynamical behavior} of polymer chains is due to the motion of the whole polymer chain rather than small motions of end beads. Moreover, one can wonder whether "nonactive" beads, which are beads that stay close to the surface during the entire course of the simulation, play a role with regards to the adsorption of polymer chains on the silica surface. To shed light on this point, we compute the number of nonactive beads, for both PAAm and PDMA, and averaged it over the five independent simulations. PAAm has 446 nonactive beads and PDMA has 427 beads. This difference is rather small and PAAm, which experimentally does not adsorbs on the silica surface, has even more nonactive beads than PDMA. Thus, nonactive beads do not explain the different behavior between PAAm and PDMA. One can see from Table \ref{tab:activity_A_hydrogel} that PAAm has more active beads than PDMA, but that PDMA's active beads undergo more events than PAAm's active beads. The consequence is that the {dynamical properties} of A beads of PAAm and A beads of PDMA is the same: the resulting frequency of the event "bead leaves" is the same for PAAm and PDMA for instance. At this stage, the study of the {dynamical properties} of A beads does not allow us to draw a conclusion about a difference in the behavior between PAAm and PDMA that would lead to different mechanical properties.

To conclude on the implicit solvent model, even if it gives a slight difference {of} $\Delta F_B$: it is more difficult for A$_{PDMA}$ beads to move away from the surface (which is confirmed by the interaction energy between A beads and surface), it does not lead to {significant} difference {of} the {dynamical properties} of PAAm and PDMA. Thus, we need a more complex system, including explicit solvent, to understand {the differing behavior between PAAm and PDMA.}
This indicates that experimentally, water not only plays a role by screening the interactions between polymer chains and surface. {Water molecules also play a steric role by competing with the adsorption of the polymers on the silica surface}; they will prevent polymer chains, or not, to adsorb on the silica surface. Therefore, water molecules have to be taken into account to explain the different behavior of PAAm and PDMA. The resulting strengthening of polymer network by the addition of nanoparticles is not only explained by energetic considerations. In the resulting strengthening of polymer network by the addition of nanoparticles, explicit solvent plays an important role.


\subsection{Solvated polymer interacting through an explicit solvent on a surface}\label{lbl:Solvated_polymer_interacting_through_explicit_solvent}

 \begin{figure}[!htbp]
  \centering
  \includegraphics[height=7cm]{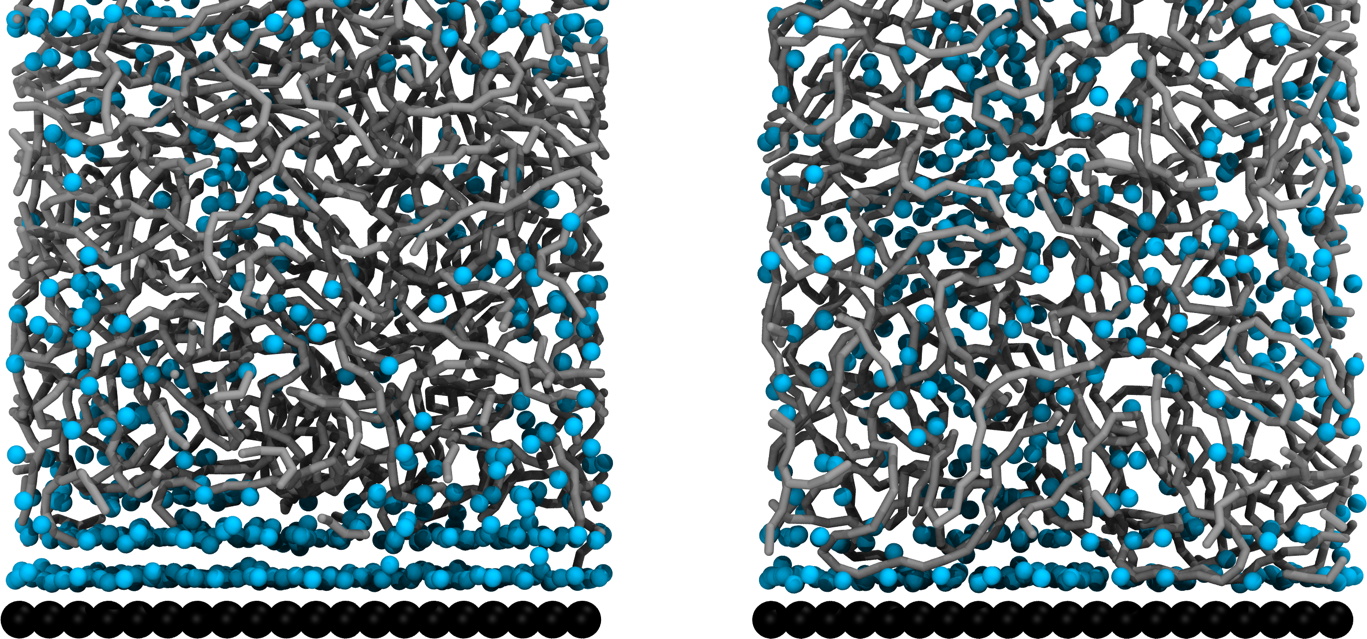}
 \caption{Snapshot of a half simulation box. Surface is in dark, solvent in blue and polymer chains are in grey. PAAm is on the right side and PDMA on the left. }
  \label{fig:hydrogel+surface}
 \end{figure}

 \begin{figure}[!htbp]
  \centering
	\includegraphics[height=5.5cm]{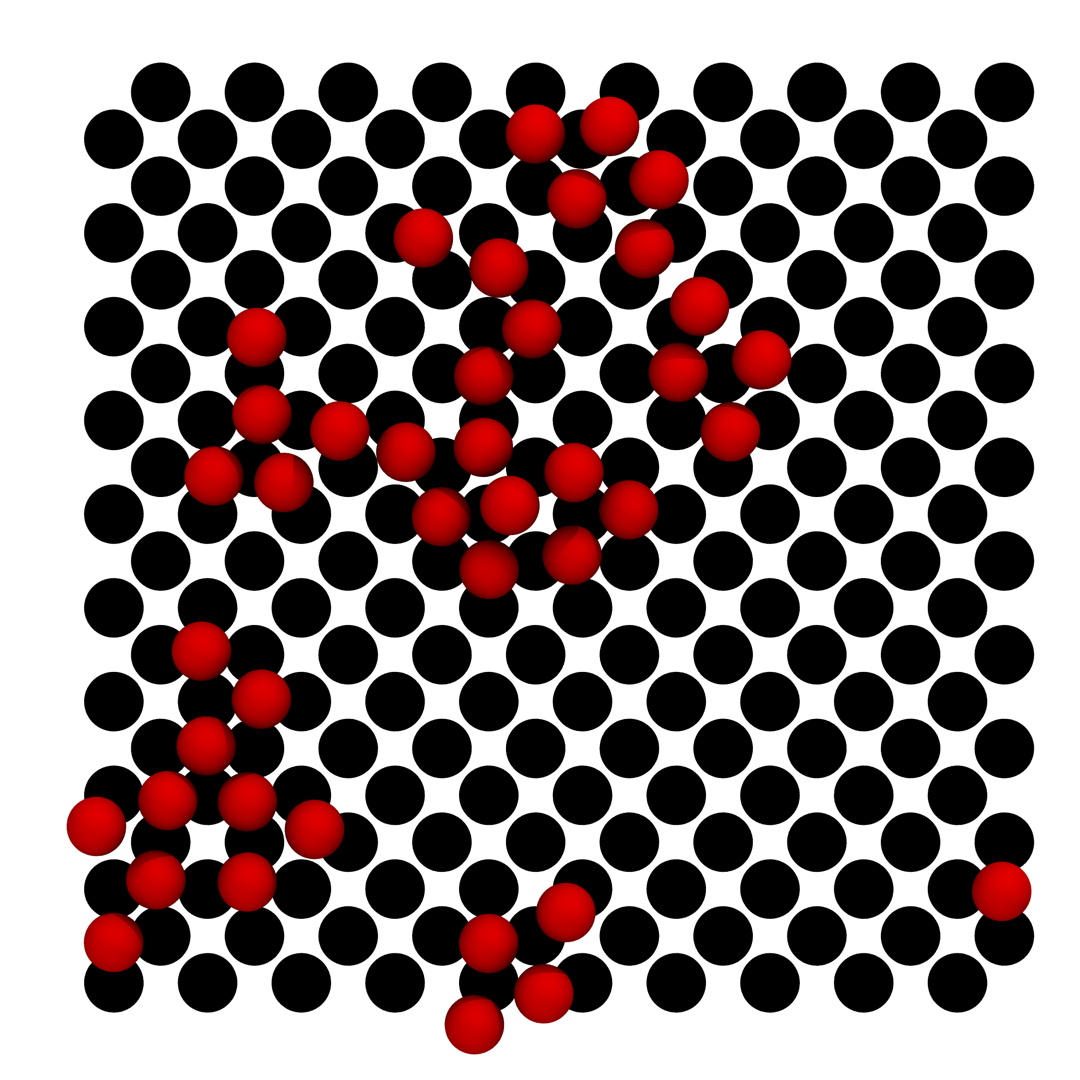}
   \caption{Top view of A$_{PDMA}$ beads (red beads) on silica surface (black beads). }
  \label{fig:PDMA+surface_top}
\end{figure}

We use systems containing 50\% of solvent {by} weight. This is a compromise between the wish to be as close to experimental {systems} as possible (which contain around 90~\% of water \cite{Rose2014}) and the computational cost that the simulation of an explicit solvent implies. Using 50\% of solvent in weight {insures having the correct number} of solvent beads in the simulation box. We use the explicit solvent described above. We still consider 24 chains containing 90 monomers {and 200 beads for the surface}. As we work with a percentage in weight, the PAAm {system} contains 960 solvent beads and PDMA {system} contains 1340 solvent beads (Fig. \ref{fig:hydrogel+surface}). {A snapshot of the simulation box is displayed on Fig. \ref{fig:snapshotPAM+solvent_nolength} of the Supplementary material. The simulation box dimensions are : $l_x$, $l_y$ = 63 \AA, $l_z$ = 80.9 \AA~for PAAm and $l_z$ = 103.4 \AA~for PDMA.} {The system was first equilibrated during 5~ns in the NPT ensemble at a pressure of one atmosphere and a temperature of 300~K, then during 5~ns in the NVT ensemble.} Figure \ref{fig:PDMA+surface_top} shows A$_{PDMA}$ beads in region I. There are fewer A beads in the vicinity of the surface than for the {polymer solution interacting through an implicit solvent} (figure \ref{fig:PDMA+surface_melt_top}). The rest of the surface is covered by solvent beads. The same behavior is observed for PAAm. The results we present are averaged over 10 independent simulations of 50~ns. The initial configuration is the same for PAAm and PDMA. First, the behavior of the polymer beads near the surface is studied, as have already been done for the polymer {solution} interacting with an implicit solvent, to see if it is modified by the presence of explicit solvent beads. Then, the way polymer beads are solvated by the explicit solvent beads is probed.

\textit{Polymer behavior with regard to the surface. }The free energy barrier $\Delta F_B$ of A beads is higher for PDMA (2.1($\pm$0.3) kcal.mol$^{-1}$) than for PAAm (1.6($\pm$0.6) kcal.mol$^{-1}$). It is more difficult for A$_{PDMA}$ beads to leave the region I. Moreover, the first two rows of Table \ref{tab:Eint_A-surface} show that in region I, A$_{PDMA}$ interact more strongly with the surface than A$_{PAAm}$. This is also what we observed with an implicit solvent. Using an explicit or an implicit solvent does not change the {ranking} of interactions between the polymers and the surface. {The strength of interaction energy with an explicit solvent is stronger than with an implicit solvent because the implicit solvent, by screening the interactions between A and S beads which are both solvophilic beads, decreases their interaction parameters.} However, what are the consequences of the use of an explicit solvent on the {residence time} of PAAm and PDMA near the surface?

\begin{table}[!htbp]\centering
\begin{tabular}{ c l c l c l c }
\hline
System & Zone I & Zone II & Zone III \\ \hline
A beads of PAAm 		&  -2.8  & -0.1 & 0   \\
A beads of PDMA		&  -4.2  & -0.3 & 0  \\
Solvent beads			&  -4.6  & -0.3 & 0  \\
\end{tabular}
\caption{Interaction Energy Between Beads and the Surface (in kcal.mol$^{-1}$).}\label{tab:Eint_A-surface}
\end{table}

\begin{table}
  \caption{Dynamic of PAAm and PDMA's A beads.}\label{tab:dynamic_A}
  \begin{tabular}{ c l c l c }
  \hline
Dynamic 									& A beads of PAAm 		& A beads of PDMA \\ \hline
Number of active beads 						&  4.1($\pm$1.7) 		& 13($\pm$5.8)   \\
Averaged number of events per active bead		&  1.5($\pm$0.6) 		& 2.5($\pm$0.7)   \\
Frequency of the event "bead leave"     			&  0.1 				& 0.4
\end{tabular}
\end{table}

One can see from Table \ref{tab:dynamic_A} that PDMA has more active beads than PAAm and that these active beads are more active than the active beads of PAAm. Not only they undergo more events, but their frequency is also higher. Therefore A beads of PDMA are slightly faster than A beads of PAAm. However, we see from table \ref{tab:Eint_A-surface} that A beads of PDMA interact more strongly with the surface than PAAm. The fact that PDMA has a stronger interaction with the surface but moves faster in the vicinity of the surface is surprising. This is certainly due to the fact that the interaction energy between A$_{PAAm}$ and the surface is low below the interaction energy of the solvent beads and the surface: PAAm is then replaced by solvent beads near the surface. This is not the case for A$_{PDMA}$, which has an interaction energy with the surface in the same range as the interaction energy between the solvent beads and the surface. Explicit solvent apparently plays a role in this system and modifies the dynamical properties of the polymer beads, even if the interaction energy ranges are not changed.

\textit{Solvation of the polymer.} To understand how the presence of solvent beads modifies the dynamics of polymer beads, it is interesting to take a close look at the way polymers are solvated by the explicit solvent. We thus analyse the solvation of  PAAm and PDMA to get some insight on a possible competition between solvent beads or polymer beads with regard to the adsorption on the surface.
We quantify the solvation of PAAm and PDMA by computing the number of first solvent neighbors of PDMA and PAAm A beads in regions I, II and III.
To do so, we count the number of solvent beads that stand at a certain distance $d_{shell}$ from A beads, where $d_{shell}$ corresponds to the radius of the first solvation shell of A beads. This procedure is done for A beads of PAAm and of PDMA and in regions I, II and III. N$_{PAAm}$ is the number of solvent first neighbors of A beads of PAAm; N$_{PDMA}$ is the number of solvent first neighbors of A beads of PDMA and we present the ratio N$_{PAAm}$/N$_{PDMA}$ in Table \ref{tab:RDF_A-solvent}.

\begin{table}[!htbp]\centering
\begin{tabular}{ c l c }
\hline
Region 		&  Ratio PDMA/PAAm \\ \hline
Full box 		&   1.4   \\
Region III		&   2.1    \\
Region II 		&   0.7   \\
Region I 		&   0.8
\end{tabular}
\caption{Ratio of First Solvent Neighbors of PDMA and PAAm.}\label{tab:RDF_A-solvent}
\end{table}

One can first note that the ratio of first neighbors of PDMA and of PAAm in the full simulation box is about 1.4, which gives the number of solvent beads for PDMA divided by the number of solvent beads for PAAm. This "full box" ratio, that we use as a reference, is compared with the ratio in regions III, II and I. In region III, the ratio goes up to 2.1, meaning that A$_{PDMA}$ has more first solvent neighbors than A$_{PAAm}$. The value of the ratio is above 1.4: there is an "excess" of solvent beads for PDMA, or a default of solvent beads for PAAm in the bulk-like region (region III), compared to the full box ratio. The tendency is well understood when we move toward the surface, in regions I and II, where the ratio lowers to 0.7 (region II) or 0.8 (region I). There is clearly, as one can see from figure \ref{fig:hydrogel+surface}, a large excess of solvent beads close to  A$_{PAAm}$ beads close to the surface (left side of the figure \ref{fig:hydrogel+surface}), compared with the box containing PDMA (right side of the figure \ref{fig:hydrogel+surface}). It is worth noting that we started the simulation of the systems containing PAAm and PDMA from the same initial configuration and led to different configurations of the solvent around the polymer. Therefore, in the box containing PAAm, solvent beads leave the bulk-like region to reach the interface with the surface resulting in a default of solvent beads in region III and an excess of solvent beads in regions I and II. In the PAAm system, solvent beads go in between the polymer chains and the surface, preventing PAAm chains from adsorbing to the surface. For PAAm, there is a competition between solvent beads and PAAm beads with regard to the adsorption on the surface. However there is also an attraction between PAAm beads and solvent beads. The attraction between PAAm and solvent is confirmed by the study of the part of the interaction energy between A beads and solvent among the interaction energy between A beads and all the other beads, which is summarized in table \ref{tab:Eint_A-solvent}. Indeed, as PAAm and PDMA do not interact with the same strength with the surrounding beads, we can not directly compare A/solvent interactions energy between PAAm and PDMA. The first row of table \ref{tab:Eint_A-solvent} shows that, in region I, the part due to A/solvent interactions among the A/everything interactions is higher for PAAm than for PDMA. Therefore A$_{PAAm}$ interacts more forcefully with solvent beads than PDMA close to the wall.

\begin{table}[!htbp]\centering
\begin{tabular}{ c | c  }
\hline
System & Part of the interaction energy in region I (\%)  \\ \hline
PAAm 			&  31.1     \\
PDMA			&  20.3    \\
\end{tabular}
\caption{Part of the Interaction Energy Between A and Solvent Beads among the Interaction Energy Between A Beads and every other Beads.}\label{tab:Eint_A-solvent}
\end{table}

To conclude on this part, we show that explicit solvent plays an important role. Even if, at first glance, the presence of explicit solvent beads does not seem to perturb the interaction between A beads and surface beads, it makes the adsorption of A$_{PAAm}$ more difficult by solvating the polymer's beads and by moving toward the interface and preventing the polymer from adsorbing on the silica surface. It is indeed easier for solvent beads to adsorb on the surface  because they are smaller and move faster than A$_{PAAm}$ beads.


\section{Conclusions and perspectives}\label{lbl:Conclusions_perspectives}

In this work, we performed coarse-grained molecular dynamic simulations of a solvated polymer near a planar model silica surface, comparing implicit and explicit solvent models. We conclude that the implicit solvent model gives reasonable results regarding the interaction energy between polymer and surface, as well as the ability for the polymer beads to move away from the surface. However, it fails to describe the dynamical properties of the polymer beads near the silica surface. The explicit solvent model, which has much higher computational cost, is necessary to describe the competition between solvent and polymer beads near the interface. In the PAAm-based system, interactions between solvent and surface are stronger than those between PAAm and surface. As PAAm is well solvated by water and prefers to be surrounded by solvent beads than close to silica beads, this prevents adhesion of the polymer on the silica surface. This is in stark contrast with the PDMA chain, where polymer--surface interactions dominate and lead to adhesion on the surface. Moreover, we highlighted the crucial role of polymer solvation for the adsorption of the polymer on the silica surface, the significant {dynamical properties} of fragments of polymer on the surface, and detail the modifications in the structure of the polymer close to the interface.

\begin{suppinfo}

The following files are available free of charge on the online Supplementary material.
\begin{itemize}
  \item Figure \ref{fig:1-11layers} is a comparison of the potential of mean force of PAAm with one and eleven layers of silica surface.
  \item Figure \ref{fig:water+surface} is a snapshot of a simulation box containing solvent beads and a silica surface.
  \item Figure \ref{fig:snapshotPAM_nolength} is a snapshot of a simulation box containing PAAm and silica surface interacting through an implicit solvent.
  \item Figure \ref{fig:active_beads} is the distribution of active beads along polymer chains.
  \item Figure \ref{fig:snapshotPAM+solvent_nolength} is a snapshot of a simulation box containing PAAm, the silica surface and an explicit solvent.

\end{itemize}

\end{suppinfo}

\begin{acknowledgement}
This work was supported by a PSL-Chimie grant and founded by the French Ministry of Higher Education and Research. The authors thank Alba Marcellan for insightful discussions. The computational resources of the Technical University of Berlin are also acknowledged. This work benefitted from access to HPC platforms provided by a GENCI grant (A0030910299).
\end{acknowledgement}

\bibliography{paper}

\end{document}